\documentstyle[twoside,epsfig]{article}

\catcode`\@=11
\long\def\@makefntext#1{
\protect\noindent \hbox to 3.2pt {\hskip-.9pt  
$^{{\eightrm\@thefnmark}}$\hfil}#1\hfill}		

\def\thefootnote{\fnsymbol{footnote}}
\def\@makefnmark{\hbox to 0pt{$^{\@thefnmark}$\hss}}	
	
\def\ps@myheadings{\let\@mkboth\@gobbletwo
\def\@oddhead{\hbox{}
\rightmark\hfil\eightrm\thepage}   
\def\@oddfoot{}\def\@evenhead{\eightrm\thepage\hfil
\leftmark\hbox{}}\def\@evenfoot{}
\def\sectionmark##1{}\def\subsectionmark##1{}}

\oddsidemargin=\evensidemargin
\renewcommand{\thefootnote}{\fnsymbol{footnote}}

\newcounter{sectionc}\newcounter{subsectionc}\newcounter{subsubsectionc}
\renewcommand{\section}[1] {\vspace{12pt}\addtocounter{sectionc}{1} 
\setcounter{subsectionc}{0}\setcounter{subsubsectionc}{0}\noindent 
	{\tenbf\thesectionc. #1}\par\vspace{5pt}}
\renewcommand{\subsection}[1] {\vspace{12pt}\addtocounter{subsectionc}{1} 
	\setcounter{subsubsectionc}{0}\noindent 
	{\bf\thesectionc.\thesubsectionc. {\kern1pt \bfit #1}}\par\vspace{5pt}}
\renewcommand{\subsubsection}[1] {\vspace{12pt}\addtocounter{subsubsectionc}{1}
	\noindent{\tenrm\thesectionc.\thesubsectionc.\thesubsubsectionc.
	{\kern1pt \tenit #1}}\par\vspace{5pt}}
\newcommand{\nonumsection}[1] {\vspace{12pt}\noindent{\tenbf #1}
	\par\vspace{5pt}}

\newcounter{appendixc}
\newcounter{subappendixc}[appendixc]
\newcounter{subsubappendixc}[subappendixc]
\renewcommand{\thesubappendixc}{\Alph{appendixc}.\arabic{subappendixc}}
\renewcommand{\thesubsubappendixc}
	{\Alph{appendixc}.\arabic{subappendixc}.\arabic{subsubappendixc}}

\renewcommand{\appendix}[1] {\vspace{12pt}
        \refstepcounter{appendixc}
        \setcounter{figure}{0}
        \setcounter{table}{0}
        \setcounter{lemma}{0}
        \setcounter{theorem}{0}
        \setcounter{corollary}{0}
        \setcounter{definition}{0}
        \setcounter{equation}{0}
        \renewcommand{\thefigure}{\Alph{appendixc}.\arabic{figure}}
        \renewcommand{\thetable}{\Alph{appendixc}.\arabic{table}}
        \renewcommand{\theappendixc}{\Alph{appendixc}}
        \renewcommand{\thelemma}{\Alph{appendixc}.\arabic{lemma}}
        \renewcommand{\thetheorem}{\Alph{appendixc}.\arabic{theorem}}
        \renewcommand{\thedefinition}{\Alph{appendixc}.\arabic{definition}}
        \renewcommand{\thecorollary}{\Alph{appendixc}.\arabic{corollary}}
        \renewcommand{\theequation}{\Alph{appendixc}.\arabic{equation}}
        \noindent{\tenbf Appendix \theappendixc #1}\par\vspace{5pt}}
\newcommand{\subappendix}[1] {\vspace{12pt}
        \refstepcounter{subappendixc}
        \noindent{\bf Appendix \thesubappendixc. {\kern1pt \bfit #1}}
	\par\vspace{5pt}}
\newcommand{\subsubappendix}[1] {\vspace{12pt}
        \refstepcounter{subsubappendixc}
        \noindent{\rm Appendix \thesubsubappendixc. {\kern1pt \tenit #1}}
	\par\vspace{5pt}}

\newcommand{\textlineskip}{\baselineskip=13pt}
\newcommand{\smalllineskip}{\baselineskip=10pt}

\def\eightcirc{
\begin{picture}(0,0)
\put(4.4,1.8){\circle{6.5}}
\end{picture}}
\def\eightcopyright{\eightcirc\kern2.7pt\hbox{\eightrm c}} 

\newcommand{\copyrightheading}[1]
	{\vspace*{-2.5cm}\smalllineskip{\flushleft
	{\footnotesize Modern Physics Letters A, #1}\\
	{\footnotesize $\eightcopyright$\, World Scientific Publishing
	 Company}\\
	 }}

\newcommand{\publisher}[2]{{\begin{center}\footnotesize\smalllineskip 
	Received #1\\
	Revised #2
	\end{center}
	}}

\def\abstracts#1#2#3{{
	\centering{\begin{minipage}{4.5in}\footnotesize\baselineskip=10pt
	\parindent=0pt #1\par 
	\parindent=15pt #2\par
	\parindent=15pt #3
	\end{minipage}}\par}} 

\newcommand{\bibit}{\nineit}
\newcommand{\bibbf}{\ninebf}
\renewenvironment{thebibliography}[1]
	{\frenchspacing
	 \ninerm\baselineskip=11pt
	 \begin{list}{\arabic{enumi}.}
        {\usecounter{enumi}\setlength{\parsep}{0pt}     
	 \setlength{\leftmargin 12.7pt}{\rightmargin 0pt} 
         \setlength{\itemsep}{0pt} \settowidth
	{\labelwidth}{#1.}\sloppy}}{\end{list}}

\newcounter{itemlistc}
\newcounter{romanlistc}
\newcounter{alphlistc}
\newcounter{arabiclistc}

\newcommand{\fcaption}[1]{
        \refstepcounter{figure}
        \setbox\@tempboxa = \hbox{\footnotesize Fig.~\thefigure. #1}
        \ifdim \wd\@tempboxa > 5in
           {\begin{center}
        \parbox{5in}{\footnotesize\smalllineskip Fig.~\thefigure. #1}
            \end{center}}
        \else
             {\begin{center}
             {\footnotesize Fig.~\thefigure. #1}
              \end{center}}
        \fi}

\newcommand{\tcaption}[1]{
        \refstepcounter{table}
        \setbox\@tempboxa = \hbox{\footnotesize Table~\thetable. #1}
        \ifdim \wd\@tempboxa > 5in
           {\begin{center}
        \parbox{5in}{\footnotesize\smalllineskip Table~\thetable. #1}
            \end{center}}
        \else
             {\begin{center}
             {\footnotesize Table~\thetable. #1}
              \end{center}}
        \fi}

\def\@citex[#1]#2{\if@filesw\immediate\write\@auxout
	{\string\citation{#2}}\fi
\def\@citea{}\@cite{\@for\@citeb:=#2\do
	{\@citea\def\@citea{,}\@ifundefined
	{b@\@citeb}{{\bf ?}\@warning
	{Citation `\@citeb' on page \thepage \space undefined}}
	{\csname b@\@citeb\endcsname}}}{#1}}

\newif\if@cghi
\def\cite{\@cghitrue\@ifnextchar [{\@tempswatrue
	\@citex}{\@tempswafalse\@citex[]}}
\def\citelow{\@cghifalse\@ifnextchar [{\@tempswatrue
	\@citex}{\@tempswafalse\@citex[]}}
\def\@cite#1#2{{$\null^{#1}$\if@tempswa\typeout
	{IJCGA warning: optional citation argument 
	ignored: `#2'} \fi}}

\def\pmb#1{\setbox0=\hbox{#1}
	\kern-.025em\copy0\kern-\wd0
	\kern.05em\copy0\kern-\wd0
	\kern-.025em\raise.0433em\box0}

\def\fnt#1#2{\footnotetext{\kern-.3em
	{$^{\mbox{\scriptsize #1}}$}{#2}}}

\def\fpage#1{\begingroup
\voffset=.3in
\thispagestyle{empty}\begin{table}[b]\centerline{\footnotesize #1}
	\end{table}\endgroup}

\def\runninghead#1#2{\pagestyle{myheadings}
\markboth{{\protect\footnotesize\it{\quad #1}}\hfill}
{\hfill{\protect\footnotesize\it{#2\quad}}}}
\font\tenrm=cmr10
\font\tenit=cmti10 
\font\tenbf=cmbx10
\font\bfit=cmbxti10 at 10pt
\font\ninerm=cmr9
\font\nineit=cmti9
\font\ninebf=cmbx9
\font\eightrm=cmr8

\def\qed{\hbox{${\vcenter{\vbox{			
   \hrule height 0.4pt\hbox{\vrule width 0.4pt height 6pt
   \kern5pt\vrule width 0.4pt}\hrule height 0.4pt}}}$}}

\renewcommand{\thefootnote}{\fnsymbol{footnote}}	

\def\sii{\qquad}

\def\b:{\begin{equation}}
\def\b:{\begin{equation}}
\def\b:{\begin{equation}}
\def\e:{\end{equation}}
\def\be:{\begin{eqnarray}}
\def\ee:{\end{eqnarray}}
\def\nn{\nonumber\\}

\def\sla{\kern -2.2mm /}

\def\bmin{\begin{minipage}{15cm} }
\def\emin{\end{minipage}}
\newcommand{\lb}[1]{\label{eqn:#1}}
\newcommand{\rf}[1]{\ref{eqn:#1}}

\newcommand{\beq}{\begin{equation}}
\newcommand{\eeq}{\end{equation}}
\newcommand{\beqa}{\begin{eqnarray}}
\newcommand{\eeqa}{\end{eqnarray}}
\newcommand{\n}{\nonumber \\}
\newcommand{\bm}[1]{\mbox{\boldmath $#1$}}
\newcommand{\pdeff}[2]{\frac{\partial #1}{\partial #2}}
\newcommand{\rec}[1]{\frac{1}{#1}}
\newcommand{\tr}{\rm tr}

\begin{document}
\setlength{\textheight}{7.7truein}  


\runninghead{Supersymmetric extension of Moyal algebra
$\ldots$}{Supersymmetric extension of Moyal algebra
$\ldots$}

\normalsize\textlineskip
\thispagestyle{empty}
\setcounter{page}{1}

\copyrightheading{}			

\vspace*{0.88truein}


\fpage{1}
\centerline{\bf Supersymmetric extension of Moyal algebra}
\baselineskip=13pt
\centerline{\bf and its application to the matrix model}
\vspace*{0.37truein}


\centerline{\footnotesize Takuya MASUDA}

\baselineskip=12pt
\centerline{\footnotesize Satoru SAITO}
\baselineskip=12pt
\centerline{\footnotesize\it Department of Physics, Tokyo Metropolitan
University}
\baselineskip=10pt
\centerline{\footnotesize\it Hachioji, Tokyo 192-0397, Japan}


\vspace*{0.225truein}

\publisher{(received date)}{(revised date)}

\vspace*{0.21truein}
\abstracts{We construct operator representation of Moyal algebra in the presence of fermionic fields. The result is used to describe the matrix model in Moyal formalism, that treat gauge degrees of freedom and outer degrees of freedom equally.
}{}{}



\vspace*{1pt}\textlineskip	
\section{Introduction}	
\vspace*{-0.5pt}
\noindent
Outer degrees of freedom can be converted into gauge degrees of
freedom through compactification. The correspondence between a commutator of matrices and a
Poisson bracket of functions used both in M-theory and I$\!$IB matrix
model implies the fact. When we use the Moyal formalism we can
interpolate these two degrees of freedom. There is, however, a problem
which we have to overcome before we apply the Moyal formalism to the
matrix model. Namely there exists no Moyal formulation of fermionic
fields, which is appropriate to describe a supersymmetric
theory. Fairlie~\cite{Fairlie2} was the first who wrote the matrix model in Moyal formalism. The
supersymmetry, however, has not been fully explored. The main purpose of this
article is to construct representation of Moyal algebra to describe the matrix model in Moyal
formalism, which treat
outer degrees of freedom and the gauge ones equally.

 Ishibashi et al. have proposed a matrix model which looks like the Green-Schwarz action of type I$\!$IB string~\cite{GS} in the Schild gauge~\cite{Schild} as a constructive definition of string theory~\cite{IKKT}. This matrix model has the manifest Lorentz invariance and ${\cal N}=2$ space-time supersymmetry.

They claim that I$\!$IB superstring theory can be regarded as a sort of
classical limit of a part of the matrix model. This correspondence is
based on the relationship between su($N$) and Poisson algebra. It is,
therefore, very interesting if we could express the matrix model lagrangian and the
Green-Schwarz action in a unified form in terms of Moyal algebra, since it
is the unique one-parameter associative deformation of the Poisson
algebra~\cite{Fletcher}. We had a problem, however, that the correspondence between
su($N$) and Poisson algebra has been shown in the case that the theory
involves only bosonic fields. On the other hand, we suggested an
operator formalism of Moyal algebra in~\cite{Kenmock}, which is a
generalization of Hamiltonian vector field. We think that the operator
formalism is suitable for a description of matrix model. Therefore it is
desirable to have an operator formalism of Moyal algebra including
fermion fields. We will
construct this in this article.

In the following two sections we review briefly the I$\!$IB Matirx model
and su($N$) $\leftrightarrow$ Poisson correspondence. We introduce our
fermionic Moyal algebra in {\S}4. Based on these arguments, we extend the matrix
model in {\S}5. This extended model has coordinates which parameterize the
world-sheet without the large N limit. Our procedure is not restricted to
the matrix model but can be applied to any system that has U($N$) gauge
invariance and supersymmetry. We will present, as an example, the Moyal
extension of the ${\cal N}=1$ SYM in {\S}6.

\setcounter{footnote}{0}
\renewcommand{\thefootnote}{\alph{footnote}}

\section{I$\!$IB matrix model}
\noindent

A large $N$ reduced model has been proposed as a nonperturbative
formulation of type IIB superstring theory\cite{IKKT}.
It is defined by the following action:
\beq
S_{\rm IKKT}  =  -{1\over g^2}{\rm Tr}\left({1\over 4}[A_{\mu},A_{\nu}][A^{\mu},A^{\nu}]
+{1\over 2}\bar{\psi}\Gamma ^{\mu}[A_{\mu},\psi ]\right) ,
\label{action}
\eeq
here $\psi$ is a ten dimensional Majorana-Weyl spinor field, and
$A_{\mu}$ and $\psi$ are $N \times N$ Hermitian matrices.
It is formulated in a manifestly covariant way which they believe
is a definite advantage over the light-cone formulation~\cite{BFSS}
to study the nonperturbative issues of superstring theory.

This action can be related to the Green-Schwarz action of superstring~\cite{GS}
by using the semiclassical correspondence in the large $N$ limit:
\beqa
-i[\;,\;] &\rightarrow& {1\over N} \{\;,\;\}_P, \n
Tr &\rightarrow& N \int {d^2 \sigma }\sqrt{\hat{g}},
\label{correspondence}
\eeqa
where $\{\ ,\ \}_P$ is a Poisson bracket.
In fact eq.(\ref{action}) reduces to the Green-Schwarz action
in the Schild gauge\cite{Schild}:
\beq
S_{\rm Schild}=\int d^2\sigma \left[\sqrt{\hat{g}}\alpha\left(
\frac{1}{4}\{X^{\mu},X^{\nu}\}_P^2
-\frac{i}{2}\bar{\psi}\Gamma^{\mu}\{X^{\mu},\psi\}_P\right)
+\beta \sqrt{\hat{g}}\right].
\label{SSchild}
\eeq
Through this correspondence, the eigenvalues of $A_{\mu}$ matrices are
identified with the space-time coordinates $X^{\mu}(\sigma )$.
The ${\cal N}=2$ supersymmetry manifests itself in  $S_{\rm Schild}$ as~\cite{IKKT}
\beqa
\delta^{(1)}\psi &=& -\frac{1}{2}
                      \sigma_{\mu\nu}\Gamma^{\mu\nu}\epsilon _1
,\quad\left(\sigma_{\mu\nu}:=\partial_0X_\mu\partial_1X_\nu-\partial_1X_\mu\partial_0X_\nu\right)\n
\delta^{(1)} X_{\mu} &=& i\bar{\epsilon _1}\Gamma_{\mu}\psi ,
\label{SSchildsym1}
\eeqa
and
\beqa
\delta^{(2)}\psi &=& \epsilon _2 ,\n
\delta^{(2)} X_{\mu} &=& 0 .
\label{SSchildsym2}
\eeqa
The ${\cal N}=2$ supersymmetry (\ref{SSchildsym1}) and (\ref{SSchildsym2}) are
directly translated into
the symmetry of $S_{\rm IKKT}$ as
\beqa
\delta^{(1)}\psi &=& \frac{i}{2}
                     [A_{\mu},A_{\nu}]\Gamma^{\mu\nu}\epsilon
_1,\quad\left(\left[A_\mu,A_\nu\right]:=A_\mu A_\nu-A_\nu
A_\mu\right)\\
\delta^{(1)} A_{\mu} &=& i\bar{\epsilon _1}\Gamma_{\mu}\psi ,
\label{Ssym1}
\eeqa
and
\beqa
\delta^{(2)}\psi &=& \epsilon _2 ,\n
\delta^{(2)} A_{\mu} &=& 0.
\label{Ssym2}
\eeqa

\section{Algebraic background}
\noindent
We consider the algebraic background of the correspondence (\ref{correspondence}).

The bases of su($N$) algebra can be written as~\cite{Fairlie1}
\[
 J_{(m_1,m_2)}=\omega^{m_1m_2/2}g^{m_1}h^{m_2},
\]
where $g$ and $h$ are matrices
\[
 g=\left(\matrix{
1 \cr
&\omega \cr
&&\omega^2 \cr
&&&\ddots \cr
&&&&\omega^{N-1} \cr}\right),\quad h=\left(\matrix{
0&1&0&\cdots&0 \cr
0&0&1&\cdots&0 \cr
\vdots&\vdots&\vdots&\ddots&\vdots \cr
0&0&0&\cdots&1 \cr
-1&0&0&\cdots&0 \cr}\right),
\]
which satisfy
\[
 g^N=h^N=-1,\quad hg=\omega g h,\quad\omega=\exp\left(2\pi i/N\right).
\]
With these bases, su($N$) is expressed as, using the notation $\bm{m}=(m_1,m_2)$,
\begin{eqnarray}
&&\left[J_{\bm{m}},J_{\bm{n}}\right]=-2i\sin\left[\frac{\pi}{N}\left(\bm{m}\times\bm{n}\right)\right]J_{\bm{m}+\bm{n}}\label{suN},\\
&&\left(0\le m_i,n_i\le N-1,\quad \bm{m},\bm{n}\ne 0\right)\nonumber
\end{eqnarray}

On the other hand the Poisson operator,
\[
 X_f=\pdeff{f}{q}\pdeff{}{p}-\pdeff{f}{p}\pdeff{}{q}=\nabla f\times\nabla
\]
satisfies the commutation relation
\[
 \left[X_f,X_g\right]=X_{\{f,g\}_P},
\]
which can be expressed as
\begin{equation}
 \left[X_{\bm{m}},X_{\bm{n}}\right]=-(\bm{m}\times\bm{n})X_{\bm{m}+\bm{n}},\quad X_{\bm{m}}:=-ie^{-i\bm{m}\cdot\bm{q}}\bm{m}\times\nabla\label{Poisson}
\end{equation}
in Fourier components through the transformation $\displaystyle
f(\bm{q})=\sum_{\bm{m}}f_{\bm{m}}e^{-i\bm{m}\cdot\bm{q}}$. Therefore
(\ref{suN}) coincides, up to a constant factor, with (\ref{Poisson}) in the $N\to\infty$ limit. It is this accordance that underlies the correspondence (\ref{correspondence})~\cite{Fairlie1}.

\

It is well-known that Moyal bracket
\begin{eqnarray*}
 \{f,g\}_M&=&\lim_{{q'\to q\atop p'\to
p}}\rec{\lambda}\sin\left[\lambda\left(\pdeff{}{q'}\pdeff{}{p}-\pdeff{}{p'}\pdeff{}{q}\right)\right]f(q',p')g(q,p)\\
&=&\lim_{\bm{q}'\to\bm{q}}\rec{\lambda}\sin\left[\frac{}{}\!
\lambda\left(\nabla'\times\nabla\right)\right]f(\bm{q}')g(\bm{q})
\end{eqnarray*}
is the unique one-parameter associative deformation of the Poisson
bracket~\cite{Fletcher}, and the algebra (\ref{Poisson}) is modified into~\cite{Fairlie1}
\begin{equation}
[K_{\bm{m}},K_{\bm{n}}]=\rec{\lambda}\sin\left[\frac{}{}\!
 \lambda(\bm{m}\times\bm{n})\right]K_{\bm{m}+\bm{n}},\label{K} 
\end{equation}
Thus we can see that Moyal algebra corresponds to
su($N$) when the parameter $\lambda$ is set to $\pi/N$, and to Poisson algebra
in the $\lambda\to 0$ limit.





\section{Moyal operator for a fermionic field}
\noindent
A supersymmetric extension of the algebra (\ref{K}) was discussed in~\cite{Fairlie1}
\be:
\{F'_{\bm{m}},F'_{\bm{n}}\}&=&\cos[\lambda(\bm{m}\times \bm{n})]K_{\bm{m}+\bm{n}}\nn
\left[K_{\bm{m}},F'_{\bm{n}}\right]&=&{1\over\lambda}\sin\left[\lambda(\bm{m}\times \bm{n})\right]F'_{\bm{m}+\bm{n}}
\lb{super Moyal modoki}
\ee:
They are realized by
\b:
K_{\bm{m}}={1\over i\lambda}F'_{\bm{m}}={1\over 2i\lambda}e^{i(2\lambda m_1\hat p+m_2\hat x)},\sii [\hat x, \hat p]=i\lambda
\e:
as well as by
\b:
K_{\bm{m}}=-{1\over \lambda}F'_{\bm{m}}={i\over 2\lambda}e^{i\bm{m}\bm{q}}\exp[-\lambda(\bm{m}\times\nabla)].
\e:

We want to generalize this superalgebra to include fields. It should be done by using the basis-independent differential operator realization $K_f$, which was introduced in~\cite{Fairlie1}:
\b:
K_f:={1\over 2i\lambda}f\left(x+i\lambda{\partial\over\partial p},\ p-i\lambda{\partial\over\partial x}\right).
\lb{K_f}
\e:
However, a problem arises if we incorporate a fermionic field in a similar way and use the algebra $(\rf{super Moyal modoki})$. In $(\rf{super Moyal modoki})$ the effect of statistics has been taken into account without reference to fields. 

There is an alternative realization of the Moyal bracket for bosonic fields. We have proposed in~\cite{Kenmock} a deformation of Hamilton vector fields which provides Moyal bracket in the place of Poisson bracket. It is a little modification of $(\rf{K_f})$, which we denote as $B_f$:
\begin{eqnarray}
B_f&:=&\lim_{{q'\to q\atop p'\to p}}\frac{i}{\lambda}\sin\left[\lambda\left(\pdeff{}{q'}\pdeff{}{p}-\pdeff{}{p'}\pdeff{}{q}\right)\right]f(q',p')\\
&=&\lim_{\bm{q}'\to\bm{q}}\frac{i}{\lambda}\sin\left[\frac{}{}\! \lambda\left(\nabla'\times\nabla\right)\right]f(\bm{q}')
\end{eqnarray}

We can check that the commutation relation among the operators
\b:
[B_f,\ B_g]=B_{i\{f,g\}_M}
\lb{boson Moyal}
\e:
holds. There must be a fermionic counterpart of this operator in order to have a supersymmetric algebra. To this end we introduce the following `operator'
\begin{eqnarray}
F_\psi&:=&-\lim_{{q'\to q\atop p'\to
p}}\cos\left[\lambda\left(\pdeff{}{q'}\pdeff{}{p}-\pdeff{}{p'}\pdeff{}{q}\right)\right]\psi(q',p')\\
&=&-\lim_{\bm{q}'\to\bm{q}}\cos\left[\lambda\left(\frac{}{}\! \nabla'\times\nabla\right)\right]\psi(\bm{q}')
\end{eqnarray}
associated with a fermionic field $\psi$. Then we are ready to convince ourselves that the commutation relations
\be:
\{F_\psi,\ F_\chi\}&=&\lambda^2 B_{i\{\psi,\chi\}_M}\nn
\left[B_{f},\ F_{\psi}\right] &=& F_{i\{f,\psi\}_M}
\lb{super Moyal}
\ee:
are correct, when the statistics of the fields are considered, where the
Moyal bracket of fermions is defined as
\begin{eqnarray*}
 \{\psi,\chi\}_M&=&lim_{{q'\to q\atop p'\to
p}}\rec{\lambda}\sin\left[\lambda\left(\pdeff{}{q}\pdeff{}{p'}-\pdeff{}{p}\pdeff{}{q'}\right)\right]\psi(q,p)\chi(q',p')\\
&=&\lim_{\bm{q}'\to\bm{q}}\rec{\lambda}\sin\left[\frac{}{}\!
					    \lambda\left(\nabla\times\nabla'\right)\right]\psi(\bm{q})\chi(\bm{q}').
\end{eqnarray*}

Moreover we can see that behind this superalgebra $(\rf{boson Moyal},\ \rf{super Moyal})$ there exists an algebra of generators which admit the su$(N)$ reduction. To show this we write the operators in Fourier components:
\be:
B_f&=&\sum_{\bm{m}}f_{\bm{m}}B_{\bm{m}},\sii\left(f(\bm{q})=\sum_{\bm{m}}f_{\bm{m}}e^{i\bm{m}\bm{q}}\right),\nn
F_\psi&=&\sum_{\bm{m}}\psi_{\bm{m}}F_{\bm{m}},\sii\left(\psi(\bm{q})=\sum_{\bm{m}}\psi_{\bm{m}}e^{i\bm{m}\bm{q}}\right),
\ee:
where
\be:
B_{\bm{m}}&:=&{1\over\lambda}e^{-i\bm{m}\bm{q}}\sinh[\lambda(\bm{m}\times\nabla)]\nn
F_{\bm{m}}&:=&-e^{-i\bm{m}\bm{q}}\cosh[\lambda(\bm{m}\times\nabla)].
\ee:
Then the generators satisfy a closed algebra
\be:
\left[B_{\bm{m}},\ B_{\bm{n}}\right]&=&-{i\over\lambda}\sin\left[\lambda(\bm{m}\times\bm{n})\right]B_{\bm{m}+\bm{n}}\nn
\left[F_{\bm{m}},\ F_{\bm{n}}\right]&=&-{i\lambda}\sin\left[\lambda(\bm{m}\times\bm{n})\right]B_{\bm{m}+\bm{n}}\nn
\left[B_{\bm{m}},\ F_{\bm{n}}\right]&=&-{i\over\lambda}\sin\left[\lambda(\bm{m}\times\bm{n})\right]F_{\bm{m}+\bm{n}}
\ee:
The structure constants of these commutators are not only all common,
but also agree with one of (11). Therefore the reduction to the su$(N)$
algebra still remains supersymmetric. We like to emphasize the
apparent difference of our commutators from those of (12). The
anticommutation relation between fermionic operators in (21) arises due
to the statistics of their fields.


\section{Moyal operator formalism in the matrix model}
\noindent
Based on the arguments presented in the previous sections, we can express the matrix model lagrangian and the Green-Schwarz action in a unified form in Moyal formalism as

\beq
 S=-\rec{g^2}\tr\left(\rec{4}[B_{X_\mu},B_{X_\nu}][B_{X^\mu},B_{X^\nu}]+\rec{2}F_{\bar{\psi}}\Gamma^\mu\left[B_{X_\mu},F_{\psi}\right]\right)\label{MMM}
\eeq
where $\left[\quad ,\quad\right]$ is a commutator of operators and not
of matrices, and ``$\tr$'' denotes the integration over the world sheet
parameters and the sum over the complete set
of functions on the world sheet.
This action is invariant under the ${\cal N}=2$ transformations
\beqa
\delta^{(1)}F_{\psi}&=& -\frac{1}{2}
                      \Sigma_{\mu\nu}\Gamma^{\mu\nu}\epsilon _1
,\quad\left(\Sigma_{\mu\nu}=\partial_0B_{X_\mu}\partial_1B_{X_\nu}-\partial_1B_{X_\mu}\partial_0B_{X_\nu}\right)\n
\delta^{(1)} B_{X_{\mu}} &=& i\bar{\epsilon _1}\Gamma_{\mu}F_{\psi} ,
\eeqa
and
\beqa
\delta^{(2)}F_{\psi} &=& \epsilon _2 ,\n
\delta^{(2)}B_{X_{\mu}} &=& 0 .
\eeqa
It already has coordinates to parameterize the world-sheet without the
$\lambda\to 0$ limit.

We would like to mention the difference of our approach from one by Fairlie \cite{Fairlie2}. The action proposed in this theory is
\begin{eqnarray*} 
S_{\rm Fairlie}&=&\rec{2\pi\alpha'}\int d\alpha d\beta d\sigma d\tau\left(\frac{}{}(D_\mu X)^2+\cos\left\{\theta^T,\not{D}\theta\right\}+g_s^2\tr F_{\mu\nu}^2\right.\\
&-&\left.\left({1\over\lambda g_s^2}\sin\{X^\mu,X^\nu\}\right)^2+{1\over g_s}\cos\left\{\psi^T\Gamma_\mu,\ {1\over\lambda}\sin\{X^\mu,\psi\}\right\}\right)
\end{eqnarray*}
here the phase space variables $\alpha, \beta$ are introduced to
parametrise the fields instead of matrix indices. In this theory the
commutators of matrices are deformed, while we deform the matrices
themselves and leave the operation of product as the same as the
operation of matrices. Moreover the action (\ref{MMM}) has manifest
${\cal N}=2$ supersymmetry.

\section{Conclusion}
\noindent
We considered the algebraic relationship behind the correspondence
between I$\!$IB matrix model and the Green-Schwarz action and noticed
that the both can be described in a unified form in Moyal formalism if
the correspondence can be shown even in the presence of fermionic
fields. We have shown this correspondence by constructing a closed
algebra. Although the extension is originally motivated by the matrix model, our procedure is not restricted to that case but can be applied to any
system that has both U($N$) gauge invariance and supersymmetry. As an example,
we present a Moyal extension of the ${\cal N}=1$ SYM.

The ${\cal N}=1$ SYM lagrangian density
\[
 {\cal L}_{\rm SYM}=-\rec{4}F_{\mu\nu}^aF^{\mu\nu a}-i\lambda^{\dagger
a}\bar{\sigma}^\mu D_\mu\lambda^a+\rec{2}D^aD^a
\]
can be extended to the lagrangian density ${\cal L}_{\rm M}$, where
\begin{eqnarray}
 {\cal L}_M&=&-\tr\left(\rec{4}Y_{\mu\nu}Y^{\mu\nu}+i F_{\lambda^\dagger}\bar{\sigma}^\mu D_\mu F_{\lambda}-\rec{2}B_D B_D-\frac{i}{8}\epsilon^{\mu\nu\rho\sigma}Y_{\mu\nu}Y_{\rho\sigma}\right)\label{SYMoyal}\\
 Y_{\mu\nu}&=&\partial_\mu B_{A_\nu}-\partial_\nu B_{A_\mu}+ig\left[B_{A_\mu},B_{A_\nu}\right]
\end{eqnarray}
and
\[
 D_\mu F_{\lambda}=\partial_\mu F_{\lambda}+ig\left[B_{A_\mu},F_{\lambda}\right].
\]
This lagrangian is invariant under supertransformations
\begin{eqnarray}
\delta B_{A_\mu}&=&-\rec{\sqrt{2}}\left[\epsilon^\dagger\bar{\sigma}_\mu F_{\lambda}+F_{\lambda^\dagger}\bar{\sigma}_\mu\epsilon\right], \\
\delta F_{\lambda_\alpha}&=&-\frac{i}{2\sqrt{2}}(\sigma^\rho\bar{\sigma}^\sigma\epsilon)_\alpha Y_{\rho\sigma}+\rec{\sqrt{2}}\epsilon_\alpha B_D,\\
\delta B_D&=&\frac{i}{\sqrt{2}}\left[\epsilon^\dagger\bar{\sigma}^\mu D_\mu F_{\lambda}-D_\mu F_{\lambda^\dagger}\bar{\sigma}^\mu\epsilon\right]
\end{eqnarray}
as well as ``gauge'' transformations 
\begin{eqnarray*}
\delta_M B_{A_\mu}&=&-\partial_\mu F_{\Lambda}-ig\left[B_{A_\mu},B_{\Lambda}\right]\\
\delta_M F_{\lambda}&=&-ig\left[F_{\lambda},B_{\Lambda}\right],
\end{eqnarray*}
where $\Lambda$ is a transformation parameter.

When we describe $S_{\rm IKKT}$ in Moyal formalism new degrees of
       freedom appear. How should we interpret these ? It gives a clue to
       remember $S_{\rm IKKT}$ is originally a low energy effective
       action of $N$ coincident $(-1)$-branes. The theory with a $p$-brane
       compactified in a direction perpendicular to the brane is
       T-dual to the theory with a $(p+1)$-brane compactified in a
       direction parallel to the brane. So the theory
       with $(-1)$-branes compactified twice in directions perpendicular to the
       branes is T-dual to the theory with 1-branes compactifed in
       directions parallel to the branes. Thus, in dual picture, there
       are two parameters to parameterize the 1-branes. The two
       parameters which appear when we describe the matrix model in
       Moyal formalism indicate that gauge degrees of freedom are
       obtained from outer degrees of freedom through compactification
       and that Moyal formalism treats these two degrees of freedom
equally.


\nonumsection{Acknowledgments}
\noindent
The authors would like to thank Shigeo Fujita and Nobuki Suzuki for
their collaboration in the beginning part of this work.
One of the authors, S.S. would like to thank Professor D.Fairlie for
interesting discussions.
This work is supported in part by the Grant-in-Aid for General
Scientific Research from the Ministry of Education, Science, Sports and
Culture, Japan (no 1060278). 



\begin{thebibliography}{000}

%
%
%
%

\bibitem{Fairlie2}
David B. Fairlie, {\bibit Mod. Phys. Lett.}
{\bibbf A13}, 263 (1998).

\bibitem{GS}
M. Green and J. Schwarz, {\bibit Phys. Lett.}
{\bibbf 136B}, 367 (1984).

\bibitem{Schild}
A. Schild, {\bibit Phys. Rev.}
{\bibbf D16}, 1722 (1977).

\bibitem{IKKT}
N. Ishibashi, H. Kawai, Y. Kitazawa and A. Tsuchiya, {\bibit
 Nucl. Phys.}
{\bibbf B498}, 467 (1997).

\bibitem{Fletcher}
P. Fletcher, {\bibit Phys. Lett.}
{\bibbf B248}, 323 (1990).

\bibitem{Kenmock}
R. Kemmoku and S. Saito, {\bibit J. Phys. Soc. Jpn.}
{\bibbf 65}, 1881 (1996).

\bibitem{BFSS}
T. Banks, W. Fischler, S.H. Shenker and L. Susskind, {\bibit Phys. Rev.}
{\bibbf D55}, 5112 (1997).

\bibitem{Fairlie1}
D. Fairlie, P. Fletcher and C. Zachos, {\bibit J. Math. Phys.}
{\bibbf 31}, 1088 (1990).

F. Bayen, M. Flato, A. Lichnerowicz and D. Sternheimer, {\bibit
 Ann. Phys.}
{\bibbf 111}, 61\&111 (1978).

E. G. Floratos, {\bibit Phys. Lett. B}
{\bf 228}, 335 (1989).
\end{thebibliography}
\end{document}